\documentclass[english,aip,amsmath,amssymb,reprint,jcp]{revtex4-2}
\usepackage[T1]{fontenc}
\usepackage[latin9]{inputenc}
\usepackage{color}
\usepackage{array}
\usepackage{float}
\usepackage{booktabs}
\usepackage{units}
\usepackage{multirow}
\usepackage{amsmath}
\usepackage{graphicx}

\makeatletter

\newcommand*\LyXPunctSpace{\hphantom{,}}
\providecommand{\tabularnewline}{\\}

\providecommand*{\code}[1]{\texttt{#1}}
\providecommand*{\strong}[1]{\textbf{#1}}

\makeatother

\usepackage{babel}
\begin{document}
\title{Development of an Optimized Parameter Set for Monovalent Ions in the
Reference Interaction Site Model of Solvation}
\author{Felipe Silva Carvalho}
\affiliation{Department of Physics and Astronomy, California State University,
Northridge, Northridge, CA 91330}
\altaffiliation[Current address:]{ Division of Chemistry and Chemical Engineering, California Institute of Technology, Pasadena, CA 91125}

\author{Alexander McMahon}
\affiliation{Department of Mathematics, California State University, Northridge,
Northridge, CA 91330}
\author{David A. Case}
\affiliation{Department of Chemistry and Chemical Biology, Rutgers University,
Piscataway, NJ 08854}
\author{Tyler Luchko}
\affiliation{Department of Physics and Astronomy, California State University,
Northridge, Northridge, CA 91330}
\email{tluchko@csun.edu}

\date{\today}
\begin{abstract}
Accurate modeling of aqueous monovalent ions is essential for understanding
the function of biomolecules, such as nucleic acid stability and binding
of charged drugs to protein targets. The 1D and 3D reference interaction
site models (1D- and 3D-RISM) of molecular solvation, as implemented
in the AmberTools molecular modeling suite, are well suited for modeling
mixtures of ionic species around biomolecules across a wide range
of concentrations. However, the available ion model parameters were
optimized for molecular dynamics simulations, not for the RISM framework,
which includes a closure approximation. To address this, we optimized
the Lennard-Jones 12-6 model for monovalent ions for 1D-RISM with
the partial series expansion of order 3 closure by fitting to experimental
values of ion-oxygen distance (IOD), hydration free energy (HFE),
partial molar volume (PMV) and mean activity coefficient. The new
parameter set demonstrated significant improvement in HFE, IOD, and
mean activity coefficients, whereas no overall change was observed
for the PMV. A second optimization step, introducing non-bonded fix
(NBFIX) parameters into the model, was necessary to account for the
cation-anion interactions that affect the mean activity coefficients.
The new parameters were validated at finite salt concentrations against
experimental data for 16 ion pairs and showed improved accuracy for
12 of them, with predictions for CsI and LiI ranked second best, while
those for CsF and LiCl ranked third best among the tested parameter
sets. The isothermal compressibilities for NaCl, KCl and LiCl were
compared against experimental data. Although 1D-RISM overestimated
the value for pure water in approximately 40\%, the relative change
as a function of salt concentration was improved with the new parameter
set for NaCl and KCl. 1D-RISM results obtained with the new NaCl parameters
were used to calculate the preferential interaction parameter of the
ions around the 24L B-DNA using 3D-RISM. The new parameters demonstrated
better agreement with experiment at physiological and higher concentrations.
At lower concentrations, the results primarily depended on the closure
with little effect from the ion parameters. Overall, the ion parameters
specifically developed for RISM show improved accuracy at infinite
dilution and finite concentrations. No difference was observed for
the preferential interaction parameters and isothermal compressibility
calculations when comparing NBFIX and non-NBFIX parameters. However,
the NBFIX parameters are numerically more stable at higher concentrations.
\end{abstract}
\maketitle

\section{Introduction}

Monovalent ions play important roles both in biology and technology.
Sodium and potassium cations are important to signal transduction
in neurons, maintain physiological levels of osmotic pressure and
electrochemical potential in cells \citep{matsarskaia2020multivalent,squire2009encyclopedia}
and modulate the reaction rate of RNA cleavage by affecting the binding
of divalent ions to deoxyribozymes-RNA complexes.\citep{rosenbach2020influence}
Monovalent ions, such as F\textsuperscript{-}, Cl\textsuperscript{-},
Na\textsuperscript{+} and K\textsuperscript{+}, have been investigated
for the development of aqueous batteries, as a more environmentally
friendly, cheaper, and safer alternative to the existing lithium batteries.
\citep{chao2020roadmap,bin2018progress,pasta2012high} Thus, accurately
modeling those ions in solution is crucial for acquiring realistic
results from simulations.

Molecular models of such systems generally use either an explicit
solvent method, in which all solute and solvent atoms are considered,
or implicit solvent method, in which the solvent's effect on the solute
are acquired through a set of approximations. Implicit solvent methods
are computationally cheaper than the explicit solvents and some common
approximations are Poisson-Boltzmann (PB), generalized Born (GB) and
integral equation theories (IET). However, the PB and GB oversimplify
the physics of solvation by representing the water as a featureless
dielectric continuum and neglecting the ion correlations.\citep{silva2022investigation}
On the other hand, IETs, such as the reference site interaction model
(RISM) \citep{pratt1977interaction,chandler1972optimized}, solve
for the density distributions of explicit solvent models, which are
the same models used in molecular dynamics (MD) simulations. For aqueous
salt solutions, typical models include SPC/E \citep{berendsen1987themissing}
or TIP3P \citep{jorgensen1983comparison} for water and Li-Merz (LM)\citep{li2015systematic,li2013rational}
and Joung-Cheatham (JC) \citep{joung2008determination} for monovalent
ions. For example, JC parameters were used to calculate preferential
interaction parameters of monovalent ions around 24L B-DNA showing
good agreement with experimental data.\citep{giambacsu2014ion,giambacsu2015competitive}

The JC and LM parameter sets are available in AmberTools molecular
modeling suite\citep{case2023ambertools} for both MD and RISM calculations.
However, Fedotova and Kruchinin\citep{fedotova2011calculations} have
observed that the performance of RISM for hydration free energy calculations
may depend significantly on the ionic parameter sets used. Furthermore,
RISM theory contains approximations that may result in distortions
of the solvent structure and thermodynamics. For aqueous ionic solutions
at infinite dilution, inaccuracies may arise in the hydration free
energy (HFE), activity, ion-oxygen distance (IOD), or the partial
molar volume (PMV) of the ions. Also, deviations in the mean activity
coefficients may be observed as the salt concentration increases and
the system is no longer at the ideal dilute condition. Modeling the
electrolyte solution with 1D-RISM is the first step while studying,
for example, the ionic environment around proteins and nucleic acids.
Therefore, one needs an accurate description of the solvent within
the theoretical framework in order to produce better predictions.

In this work, we present a set of monovalent ion Lennard-Jones parameters
optimized specifically for RISM calculations, with the aim of improving
properties calculated both at infinite dilution and finite concentrations.
Such specific optimizations have been carried out for molecular liquids,
such as hydrogen chloride\citep{gutwerk1994determination}, ethyliodine\citep{bertagnolli1991determination}
and cyclohexane\citep{luchko2016sampl5}. Here, we have solved the
1D-RISM equations for a range of $\epsilon$ and $\sigma$ values,
which includes the parameter values for all ions currently available
on AmberTools. Then, comparing the results against experimental HFE,
IOD and PMV data,\citep{li2015systematic,couture1956partial} it was
possible to determine the final parameters for RISM at infinite dilution.
As the cation-anion interactions were not taken into account in the
first optimization step, non-bonded fix (NBFIX) exceptions to the
standard mixing rules were parameterized for these interactions. The
results were compared with those obtained using the current parameter
sets implemented in AmberTools, and an overall improvement was observed
with the new RISM-optimized parameters. Although the NBFIX parameters
for monovalent salts are unnecessary for most 3D-RISM calculations,
they may offer improved numerical stability at higher salt concentrations.

\section{Theoretical background}

\subsection{The 1D-RISM equations}

In order to study how the solvent, \emph{e.g.}, an electrolyte solution,
is structured around a solute, \emph{e.g.}, a biomolecule, one needs
to solve the 3D-RISM equations. The first step in this process is
to obtain a model of the bulk solvent, which is acquired from the
solution of the dielectrically consistent 1D-RISM equation (DRISM)\citep{perkyns1992dielectrically}
\begin{align}
h'_{\alpha\beta}(r) & =\sum_{\nu}^{N_{\text{site}}}\sum_{\gamma}^{N_{\text{site}}}\omega'_{\alpha\nu}(r)*C_{\nu\gamma}(r)*\omega'_{\gamma\beta}(r)\nonumber \\
 & \phantom{=\sum_{\nu}^{N_{\text{site}}}\sum_{\gamma}^{N_{\text{site}}}}+\omega'_{\alpha\nu}(r)*C_{\nu\gamma}(r)*\rho_{\gamma}h'_{\gamma\beta}(r),\label{eq:1DRISMeq-1}
\end{align}
where $r$ is the separation between two sites, $h'(r)$ is the modified
total correlation function, $C(r)$ is the direct correlation function,
$\omega\left(r\right)$ is the modified intramolecular correlation
function, $\rho$ is the bulk number density, and subscripts indicate
the label of interacting sites, spanning all solvent sites. Solutes
at infinite dilution are included as solvent sites with zero density.
DRISM imposes dielectric consistency for water and ions through $h'=h-\zeta$
and $\omega'=\omega+\zeta$, where $h$ and $\omega$ are the usual
total and intramolecular correlation functions and $\zeta$ is calculated
based on the desired medium's dielectric constant.

As Equation (\ref{eq:1DRISMeq-1}) has two unknowns, a closure relation
is required to obtain a unique solution. While the closure relation
can be expressed analytically, in practice it must be approximated.\citep{hansen2006theoryof}
The hypernetted-chain (HNC) approximation\citep{morita1958theory,vanleeuwen1959new}
is known to perform well for systems with long-ranged forces.\citep{lee2021molecular}
However, it can be difficult to converge solutions using this approximation
due to the strong nonlinearities. Kovalenko and Hirata proposed\citep{kovalenko1999self-consistent}
a partially linearized version of HNC (KH), which is easier to converge,
and later S. M. \LyXPunctSpace Kast and T. \LyXPunctSpace Kloss\citep{kast2008closed-form}
generalized this approximation as the partial series expansion of
order-$n$ (PSE-n),
\begin{align}
h_{\alpha\beta}(r) & =\begin{cases}
e^{t_{\alpha\beta}^{*}(r)}-1 & \text{for }t^{*}(r)\leq0\\
\left(\sum_{i=0}^{n}\frac{\left(t_{\alpha\beta}^{*}(r)\right)^{i}}{i!}\right)-1 & \text{for }t^{*}(r)>0
\end{cases}\label{eq:PSEn-closure}\\
t_{\alpha\beta}^{*}(r) & =-\frac{U_{\alpha\beta}(r)}{k_{B}T}+h_{\alpha\beta}(r)-C_{\alpha\beta}(r),
\end{align}
where $U_{\alpha\beta}(r)$ is the potential energy function between
any two solvent sites, $k_{B}$ is Boltzmann's constant, and $T$
is the temperature. The KH and HNC closures are retrieved with $n=1$
and $n=\infty$ respectively. For a given potential energy function,
temperature and site densities, Equations \ref{eq:1DRISMeq-1} and
\ref{eq:PSEn-closure} can be solved with an iterative approach.

In this work, the potential energy is the sum of the Coulomb and Lennard-Jones
interactions. The Coulomb potential energy is given by, 
\[
U_{\alpha\beta}^{\text{C}}\left(r\right)=k\frac{q_{\alpha}q_{\beta}}{r},
\]
where $k$ is the Coulomb constant and $q_{\alpha}$ is the charge
of site $\alpha$, while the Lennard-Jones potential energy is given
by
\begin{align*}
U_{\alpha\beta}^{\text{LJ}}(r) & =\sqrt{\epsilon_{\alpha}\epsilon_{\beta}}\left[\left(\frac{R_{\text{min},\alpha}+R_{\text{min},\beta}}{2r}\right)^{12}\right.\\
 & \phantom{=\sqrt{\epsilon_{\alpha}\epsilon_{\beta}}}\left.-2\left(\frac{R_{\text{min},\alpha}+R_{\text{min},\beta}}{2r}\right)^{6}\right]\\
 & =\frac{A_{\alpha\beta}}{r^{12}}-\frac{B_{\alpha\beta}}{r_{\alpha\beta}^{6}},
\end{align*}
where $R_{\text{min}}$ is the separation distance at which the minimum
energy, $\epsilon$, is obtained. $A$ and $B$ coefficients are calculated
as
\begin{align}
A_{\alpha\beta} & =\sqrt{\epsilon_{\alpha}\epsilon_{\beta}}\left(\frac{R_{\text{min},\alpha}+R_{\text{min},\beta}}{2}\right)^{12}\label{eq:LJ-A-coeff-1}\\
B_{\alpha\beta} & =2\sqrt{\epsilon_{\alpha}\epsilon_{\beta}}\left(\frac{R_{\text{min},\alpha}+R_{\text{min},\beta}}{2}\right)^{6}.\label{eq:LJ-B-coeff-1}
\end{align}

\subsection{Thermodynamic properties}

\subsubsection{Hydration free energy}

For KH and PSE-n closure relations it is possible to acquire an analytical
form for the excess chemical potential\citep{kovalenko1999self-consistent,kast2008closed-form}.
For the PSE-n closure, one has
\begin{align*}
\mu_{\alpha}^{\text{ex}} & =4\pi k_{B}T\\
 & \phantom{=}\times\sum_{\beta}\rho_{\beta}\int_{0}^{\infty}\left[\frac{h_{\alpha\beta}^{2}}{2}-C_{\alpha\beta}-\frac{h_{\alpha\beta}C_{\alpha\beta}}{2}\right.\\
 & \phantom{=\times\sum_{\beta}\rho_{\beta}\int_{0}^{\infty}}\left.-\frac{\left(t_{\alpha\beta}^{*}\right)^{n+1}}{\left(n+1\right)!}\Theta\left(t_{\alpha\beta}^{*}\right)\right]r^{2}dr,
\end{align*}
where $\Theta\left(\cdot\right)$ is the Heaviside function. The hydration
free energy (HFE) can be defined as the free energy associated with
transferring a solute at a fixed position from an ideal gas solvent
into water at the same solvent number density. At solute infinite
dilution one has
\begin{eqnarray}
\text{HFE} & = & \mu-\mu^{\text{id}}=\mu^{\text{ex}},
\end{eqnarray}
for both the Gibbs and Helmholtz free energies\citep{ben-naimStandardThermodynamicsTransfer1978,chongThermodynamicEnsembleIndependenceSolvation2015}.
Thus, one can obtain the HFE directly from the excess chemical potential
by carrying out the calculations at infinite dilution.

It is well known that the PSE-n family of closures, which includes
KH and the hypernetted-chain equations,\citep{kast2008closed-form}
underestimates the non-polar component of the HFE and several corrections
have been developed to mitigate this issue \citep{palmer2010towards,Sergiievskyi2015Solvation,truchon2014cavity}.
In this work, we employ a modified version of the Universal Correction
\citep{palmer2010towards,johnson2016small}:
\begin{equation}
\mu^{\text{UC}}=\mu^{\text{ex}}+a\bar{V}+b,\label{eq:UC}
\end{equation}
where $\bar{V}$ is the partial molar volume of the solute (see Section
\ref{subsec:Partial-molar-volume}) and $a$ and $b$ are fit parameters\citep{johnson2016small}.

We use the correction only for infinite dilution calculations because
the non-polar contribution is known to be too positive and does not
correlate well with the true non-polar behavior. If we did not apply
this correction, our subsequent optimized parameters would try to
compensate for the large non-polar contributions.

We note that when DRISM is used with the PSE-n family of closures,
free-energy inconsistencies are introduced. However, these are small
in the temperature and density regimes we are concerned with\citep{joung2013simpleelectrolyte}.
Furthermore, when DRISM total correlation functions are used with
3D-RISM in solute-solvent calculations\citep{hirataInterionicPotentialMean1983,kovalenkoThreedimensionalDensityProfiles1998},
the thermodynamics are consistent with DRISM, since DRISM defines
the solvent properties (see Supplementary Material Section S2).

\subsubsection{Partial molar volume\label{subsec:Partial-molar-volume}}

The partial molar volume, $\bar{V}$, can be calculated from the direct
correlation function and the Kirkwood-Buff theory for liquid mixtures
as\citep{luchko2016sampl5,Kirkwood1935,imai2000theoretical}

\begin{eqnarray*}
\bar{V}_{\text{ion}} & = & k_{B}T\chi_{T}\left[1-\sum_{\alpha}^{N_{\text{site}}}4\pi\rho_{\alpha}\int r^{2}C_{\alpha}(r)dr\right],
\end{eqnarray*}
in which $\chi_{T}$ is the isothermal compressibility for the pure
solvent, which was also calculated from DRISM using the standard expression\citep{mcquarrieStatisticalMechanics2000a},
\[
\chi_{T}=\frac{\beta}{\rho-\sum_{\alpha}^{N_{\text{site}}}\sum_{\beta}^{N_{\text{site}}}\rho_{\alpha}\rho_{\beta}\hat{c}_{\alpha\beta}\left(0\right)},
\]
where $\hat{c}_{\alpha\beta}\left(0\right)$ is the value of the Fourier
transform of the direct correlation function at $k=0$ for sites $\alpha$
and $\beta$.

\subsubsection{Ion-oxygen distance}

The ion-oxygen distance can be acquired from the position of the first
peak in the total correlation function of the oxygen-ion pair.

\subsubsection{Mean activity coefficient}

The mean activity coefficient, $\gamma_{\nicefrac{+}{-}}^{m}$ defined
in the scale of molal concentration, can be calculated from\citep{joung2013simpleelectrolyte}
\begin{equation}
\nu k_{B}T\ln\left(\gamma_{\nicefrac{+}{-}}^{m}\right)=\sum_{\alpha=\{+,-\}}\nu_{\alpha}\Delta\mu_{\alpha}^{\text{ex}}+\nu k_{B}T\ln\left(\frac{\rho_{w}}{\rho_{w,\infty}}\right),\label{eq:mean-activity}
\end{equation}
where $\nu_{+}$ and $\nu_{-}$ are the cation and anion stoichiometric
coefficients, $\nu=\nu_{+}+\nu_{-}$, $\Delta\mu_{\alpha}^{\text{ex}}$
is the difference of the excess chemical potential at infinite dilution
and the target ion concentration of ion site $\alpha$, $\rho_{w}$
is the number density of water at the concentration considered, and
$\rho_{w,\infty}$ is the water number density at infinite dilution.
The derivation of this result is presented in Supplementary Material
Section S1. Because the mean activity coefficient is sensitive to
small changes in $\Delta\mu_{\alpha}^{\text{ex}}$, it is important
to include the dielectric constant of the salt solution for DRISM
calculations. We note that we do not use a correction for the excess
chemical potential here (Equation \ref{eq:UC}) because we need only
the relative change.

\section{Methods}

LJ parameter optimization was carried out in two steps. First, $\nicefrac{R_{\text{min}}}{2}$
and $\epsilon$ were optimized for each ion at infinite dilution.
These parameters are intended to be used with the standard Lorentz-Berthelot
\citep{berthelot1898melange,lorentz1881ueber} mixing rules. Then,
the LJ $\nicefrac{R_{\text{min}}}{2}$ and $\epsilon$ parameters
were held fixed for water-ion interactions, while the $A$ and $B$
coefficients for cation-anion pairs were fined-tuned using calculations
over a range of finite concentrations.

In all cases, the calculations were carried out with AmberTools molecular
modeling suite \citep{case2023ambertools,luchko2010three-dimensional}
with a grid spacing of $\unit[0.025]{\mathring{A}}$ , temperature
of $\unit[298.15]{K}$ , dielectric constant of 78.4, the coincident
SPC/E (cSPC/E) model\citep{luchko2010three-dimensional}, and residual
tolerance of $10^{-8}$. Convergence was accelerated with the modified
direct inversion of the iterative subspace (MDIIS) method \citep{kovalenko1999solution},
using a maximum of 50 solution vectors. The calculations were carried
out sequentially with KH, PSE-2 and PSE-3 closures, using the solution
from the previous closure as a starting guess to the next one. Thus,
the new parameters acquired in this work were specifically optimized
for cSPC/E water model, PSE-3 closure and the absolute proton Gibbs
energy by Y. Marcus\citep{Marcus91}. For infinite dilution calculations,
the RISM equations were solved on a grid of 16384 points and the MDIIS
step size was set to 0.5. For finite concentrations, the number of
grid points and MDIIS step size were changed to 32768 and $0.7$,
respectively.

\subsection{Optimization at infinite dilution \label{subsec:Optimization-at-infinite}}

For infinite dilution calculations, we ran separate calculations for
each ion, with an ion concentration of $\unit[0]{M}$ and water concentration
of $\unit[55.34]{M}$.

The 1D-RISM equation was solved for negatively and positively charge
particles, using a non-uniform parameter scan of $\nicefrac{R_{\text{min}}}{2}$
and $\epsilon$ values. In both cases, $\epsilon$ ranged from $10^{-5}$
to $10^{-1}$ using a logarithm scale and from $10^{-1}$ to $1$
using linear scale, while the values of $\nicefrac{R_{\text{min}}}{2}$
were set from $1$ to $4.5$ using a linear scale. The 1D-RISM calculations
provided hydration free energy, partial volume and ion-oxygen distance
values at each grid point.

For each species of ion, the cost function, $f$, at each grid point
was calculated using data from experiment,
\begin{equation}
f=w_{\text{HFE}}\times R_{\text{HFE}}+w_{\text{IOD}}\times R_{\text{IOD}}+w_{\text{PMV}}\times R_{\text{PMV}},\label{eq:cost-fun}
\end{equation}
where $R_{\text{HFE}}$, $R_{\text{IOD}}$ and $R_{\text{PMV}}$ are
the relative error for the hydration free energy, ion-oxygen distance
and partial molar volume, respectively and $w_{\text{HFE}}=2$, $w_{\text{IOD}}$
and $w_{\text{PMV}}=0.01$ are the respective weights. The weights
for each property were selected to preferentially optimize the hydration
free energies, while maintaining reasonable values for the partial
molar volume. Due to the well known issues with the HFE from 1D-RISM
with KH and PSE-3 closures, we optimized against the corrected for
HFE using Equation \ref{eq:UC} with parameters $a=\unit[-0.1185]{kcal/mol\cdot\mathring{A}^{3}}$
and $b=\unit[-0.3]{kcal/mol}$ from a previously published work\citep{johnson2016small}.
From the grid plots of the cost functions, we identified the parameters
that gave the lowest value of $f$ for each ion type, and then employed
the Nelder-Mead optimization method, implemented in the SciPy library,
\citep{virtanen2020scipy} using these parameters as initial guess
to lower the cost function further.\textcolor{black}{{} The $\epsilon$
parameter was restrained to be larger than $\unit[10^{-2}]{kcal/mol}$
for anions and $\unit[0.3]{kcal/mol}$ for cations to prevent unphysical
or poorly converging parameters.}

\subsection{Nonbond fix for finite concentrations\label{subsec:Non-bond-fix-for}}

NBFIX parameters (\textit{i.e.}, exceptions to the standard mixing
rules) were obtained for all cation-anion pairs by optimizing the
mean activity coefficients of each salt with available experimental
data \citep{haynes2014crc} over a range of concentrations starting
from infinite dilution, $\unit[0]{m}$, up to $\unit[1.0]{m}$, while
holding the water-ion parameters fixed. The concentration increments
were aligned with available experimental data \citep{haynes2014crc}
to enable the cost function calculation for each data point. The dielectric
constant for each DRISM calculation was obtained by linearly interpolating
between the value for pure water and the experimentally determined
value at 1 M \citep{giesePermittivityDielectricProton1970,marcusEvaluationStaticPermittivity2013},
see Supplementary Material Section S4.B. For each salt, the values
for $A_{\text{cation-anion}}$ and $B_{\text{cation-anion}}$ were
defined within a range extending from 15\% below to 15\% above the
values calculated from the Lorentz-Berthelot mixing rules. The only
exceptions were NaI, NaBr, KI, LiBr, and LiI, which had ranges for
$A_{\text{cation-anion}}$ extending from the calculated values for
infinite dilution down to 45\% for NaI and 35\% for the others. The
mean activity coefficient was calculated for each grid point, and
the cost function was calculated as
\begin{eqnarray}
f & = & \bar{R}_{\gamma_{\nicefrac{+}{-}}},
\end{eqnarray}
where $\bar{R}_{\gamma_{\nicefrac{+}{-}}}$ is the average of the
absolute relative errors of the mean activity coefficients over all
concentrations.

The grid searches produced continuous regions of low cost function
values in the $A_{\text{cation--anion}}$--$B_{\text{cation--anion}}$
parameter space. Rather than selecting the global minimum of the cost
function, we chose the $A$ and $B$ coefficients that exhibited the
smallest relative deviation from the corresponding Lorentz--Berthelot
values while remaining within the region of low cost (Figure \ref{fig:avg-error-mac-nacl}).
To accomplish this, for each value of $B_{\text{cation--anion}}$
we identified the value of $A_{\text{cation--anion}}$ yielding the
lowest cost function and constructed a linear interpolation of these
minima. The final NBFIX parameters were then selected as the point
along this interpolated line that minimizes 
\[
\left(\frac{A_{\text{NBFIX}}-A_{0}}{A_{0}}\right)^{2}+\left(\frac{B_{\text{NBFIX}}-B_{0}}{B_{0}}\right)^{2},
\]
where $A_{0}$ and $B_{0}$ are the Lorentz--Berthelot parameters.

For several ion pairs, DRISM calculations failed to converge for small
values of the $A$ and $B$ coefficients, particularly at high salt
concentrations. To reduce potential convergence problems in DRISM
for salt mixtures or in 3D-RISM calculations, we selected NBFIX parameter
values larger than those obtained from the interpolation-based selection
procedure (see Supplementary Material Section S4.A). In cases where
the region of low cost function values coincided with the edge of
the convergence region, the NBFIX parameters were adjusted by up to
0.02 relative units toward the infinite dilution parameters (Figure
S.3). If the difference between the infinite dilution and NBFIX parameters
was less than 0.02 relative units, the infinite dilution parameters
were used. However, for CsF and LiCl, the infinite dilution parameters
lay outside the region of convergence, and the NBFIX parameters were
instead selected 0.02 relative units farther from the infinite dilution
parameters (Figure S.4).

Because the water density depends on the salt concentration, we calculated
the input solution density for 1D-RISM calculations, in units of $\unit{\nicefrac{g}{ml}}$,
$\rho$, from \citep{krumgalz1996volumetric}
\begin{eqnarray}
\rho & = & \frac{1000+mM_{MX}}{\nicefrac{1000}{\rho_{0}}+mV_{\phi,MX}},
\end{eqnarray}
where $\rho_{0}=\unit[0.9970470]{g/ml}$ is the pure water density,
$m$ is the concentration in molality, $M_{MX}$ is the salt molar
mass, and $V_{\phi,MX}$ is the salt apparent molal volume. This last
quantity was fit against experimental data by Krumgalz, Pogorelsky
and Pitzer\citep{krumgalz1996volumetric} and its functional form
is given by:

\begin{align*}
V_{\phi,MX} & =\bar{V}_{MX}^{0}+\nu|z_{M}z_{X}|\left(\frac{A_{V}}{2b}\right)\ln\left(1+bI^{\nicefrac{1}{2}}\right)\\
 & \phantom{=}+2RT\nu_{M}\nu_{X}\left[mB_{MX}^{V}+m^{2}\nu_{M}z_{M}C_{MX}^{V}\right],
\end{align*}
where $\bar{V}_{MX}^{0}$ is the partial molal volume at infinite
dilution, $\nu_{M}$ and $\nu_{X}$ are the total number of cations
and anions generated after salt dissociation, $z_{M}$ and $z_{X}$
are the respective charges, $\nu=\nu_{M}+\nu_{X}$, $R$ is the ideal
gas constant, $T$ is the temperature, and $I$ is the ionic strength.
Values for $A_{V}$ at 298.15 K and $b$, as well as the formulas
for calculating $B_{MX}^{V}$ and $C_{MX}^{V}$, are given by Krumgalz,
Pogorelsky and Pitzer\citep{krumgalz1996volumetric}.

To match the concentrations in the experimental data in 1D-RISM, it
was necessary to convert molal units to molar units as 
\begin{eqnarray}
M & = & \frac{\rho}{\nicefrac{1}{m}+\nicefrac{M_{MX}}{1000}},
\end{eqnarray}
in which $M$ is the concentration in molar units. In this case, all
salts being 1:1 electrolytes, the concentration for each ion is the
same as given by the previous equation. Then, the water concentration
in molar units was calculated as
\begin{eqnarray}
M_{\text{water}} & = & \frac{\nicefrac{\left(100\rho-0.1M_{MX}M\right)}{18.016}}{0.1}.
\end{eqnarray}
To accommodate input of the NBFIX parameters into the \code{rism1d}
program, we updated the MDL file format. In the new format, multiple
solvent species and their nonbond fix parameters may be included in
a single file. To simplify constructing MDL files from existing force
fields, including ions that use the 12-6-4 Lennard-Jones model \citep{li2014taking},
we created the \code{generateMDL} command line utility. The parameters
present in the new file format can be easily modified using the ParmEd
package\citep{shirts2017lessons,SwailsParmEd2023}. This utility will
be available in the next release of AmberTools.

\subsection{Preferential interaction parameter calculations}

Following Giamba\c{s}u et al. \citep{giambacsu2014ion}, we calculated
the preferential interaction parameters for a 24 basepair DNA molecule
in an aqueous NaCl solution using 3D-RISM. The solvent susceptibility
was calculated for concentrations ranging from $\unit[0.01]{m}$ to
$\unit[1]{m}$ with Joung-Cheatham \citep{joung2008determination},
Li-Merz \citep{li2015systematic,li2013rational}, and the new optimized
parameters using 1D-RISM. 1D-RISM setting were the same as those used
in Section \ref{subsec:Non-bond-fix-for}, except that the grid was
set to $65536$ points and the MDIIS step size was set to $0.3$.
For 3D-RISM calculations, the buffers were set to $\unit[192]{\mathring{A}}$
for $\unit[0.005]{m}$, $\unit[144]{\mathring{A}}$ for concentrations
between $\unit[0.01]{m}$ and $\unit[0.05]{m}$ and $\unit[48]{\mathring{A}}$
for concentrations up to $\unit[1]{m}$. Additional details about
the method and buffer value choices can be found in Ref. \citep{giambacsu2014ion}.

\section{Results and Discussion}

\subsection{Parameter optimization: infinite dilution}

From the Lennard-Jones parameter scans, we produced HFE, IOD and PMV
grids for each individual ion, which we compared against values from
experiment.\citep{li2015systematic,couture1956partial} Representative
relative error grid plots of each property and the cost function for
Na\textsuperscript{+} and Cl\textsuperscript{-} ions are presented
in Figures \ref{fig:Relative-error-grid-na} and \ref{fig:Relative-error-grid-cl}
(data for all ions are described in Supplementary Material Section
S3.B and provided as CSV files). Within this range of parameters,
each quantity has a basin of values for which the cost function attains
low values, but there is no overlap of the low-value basins among
the three observables, which required us to prioritize specific observables
for optimization (Section \ref{subsec:Optimization-at-infinite}).
An added consideration was the fact that the IOD values are smaller
in magnitude than HFE and PMV values. We selected our weights to prioritize
HFE because the excess chemical potential enters subsequent finite-concentration
calculations, and gave it the largest weight of $w_{\text{HFE}}=2$.
The IOD had a weight of $w_{\text{IOD}}=0.8$, since it characterizes
structural correlations and has low magnitude values compared to the
HFE and PMV values. Lastly, the PMV, which was not used in the LM
and JC optimizations, had a weight of $w_{\text{PMV}}=0.01$ to prevent
extreme or non-physical values. We found that global minima of the
resulting cost functions for each ion were not sensitive to the specific
values of the weights we selected (Supplementary Material Section
S3.A).

\begin{figure*}
\includegraphics{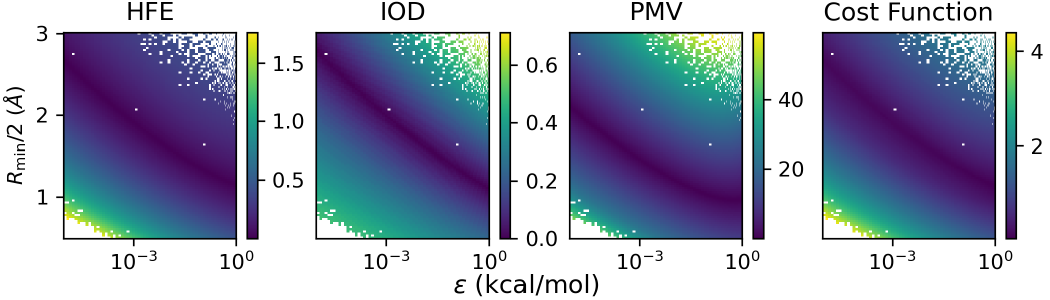}

\caption{Relative error of HFE, IOD and PMV, compared with experimental data,\citep{li2015systematic,couture1956partial}
and total cost function for sodium ion Lennard-Jones parameters. Parameters
for which the solution of 1D-RISM equations did not converge are colored
white.\label{fig:Relative-error-grid-na}}
\end{figure*}

\begin{figure*}
\includegraphics{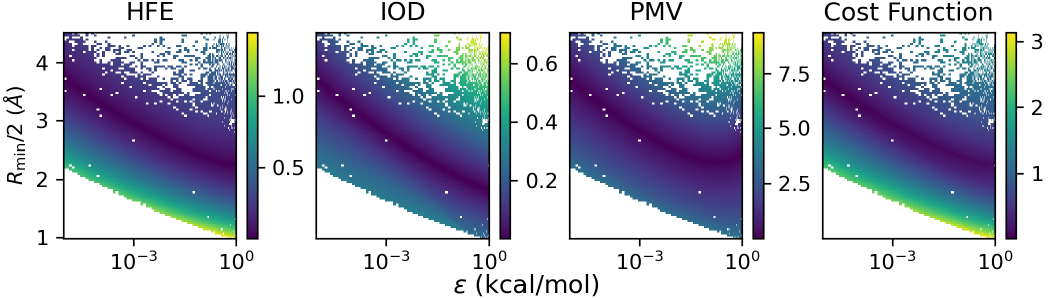}

\caption{Relative error of HFE, IOD and PMV, compared with experimental data,\citep{li2015systematic,couture1956partial}
and total cost function for chloride ion Lennard-Jones parameters.
Parameters for which the solution of 1D-RISM equations did not converge
are colored white. \label{fig:Relative-error-grid-cl}}
\end{figure*}

As with the HFE, IOD and PMV, the resulting cost function displays
low-cost bands that cover a wide region of parameter space. Variation
along the low-cost band appears small, however clear global minimum
values for both anions and cations are well defined and insensitive
to variation in the weights (Supplementary Material Section S3.A).
In the case of anions, the global minimum values lie on the far left
side of the plot. Since it represents unphysically low values of $\epsilon$,
a lower boundary for those values was defined as $10^{-2}$. For the
cations, we found that values of $\epsilon<0.3$ kcal/mol lead to
convergence issues for high ion concentrations calculations. Therefore,
we defined a lower boundary value of $0.3$ kcal/mol to ensure reliable
behavior for finite concentration calculations.

\begin{figure*}
\includegraphics{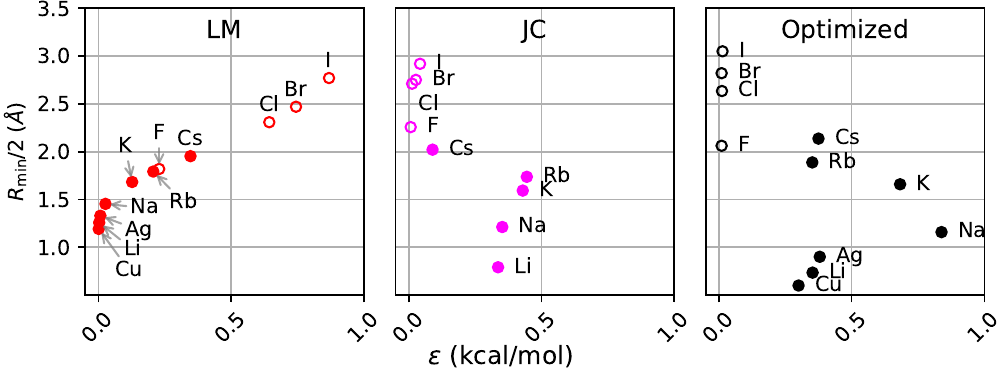}

\caption{Lennard-Jones parameter plots for all ions. LM: Li-Merz optimized
for hydration free energies \citep{li2015systematic}; JC: Joung-Cheatham
\citep{joung2008determination}; Optimized: values calculated in this
work.\label{fig:Parameter-plots}}
\end{figure*}

Our optimized $\nicefrac{R_{\text{min}}}{2}$ and $\epsilon$ values
almost completely agree with the LM and JC parameter sets for the
$\nicefrac{R_{\text{min}}}{2}$ ordering of the ions, but there is
a significant qualitative difference between the LM and JC parameter
sets for the $\epsilon$ values (Table \ref{tab:parameters}, Figure
\ref{fig:Parameter-plots}). \textcolor{black}{Note that there are
multiple LM parameter sets for monovalent ions: LJ 12-6 parameters
optimized for HFE (LM-HFE), LJ 12-6 parameters optimized for IOD (LM-IOD)
and the LJ 12-6-4 parameters (LM 12-6-4).\citep{li2015systematic}}
With exception of F, anions in the LM parameter set have larger $\epsilon$
and $\nicefrac{R_{\text{min}}}{2}$ values than cations. Conversely,
in the JC parameter set anions tend to smaller $\epsilon$ and larger
$\nicefrac{R_{\text{min}}}{2}$ than cations, with exception of Cs\textsuperscript{+}.

O\textcolor{black}{verall, o}ur optimized parameters are most consistent
with the JC parameters, for which anions have smaller values of $\epsilon$
than cations. In addition, $\nicefrac{R_{\text{min}}}{2}$ values
for the cations are close to those from JC parameter set. The main
differences are that Cs\textsuperscript{+} has a larger value of
$\nicefrac{R_{\text{min}}}{2}$ than F\textsuperscript{-} and Cs\textsuperscript{+},
Na\textsuperscript{+} and K\textsuperscript{+} have significantly
larger values of $\epsilon$ in our parameter set.

\begin{table}[H]
\caption{Monovalent ions parameters optimized for infinite dilution.\label{tab:parameters}}

\centering{}%
\begin{tabular}{ccc}
\toprule 
Ion & $\epsilon$ (kcal/mol) & $\nicefrac{R_{min}}{2}$ (Å)\tabularnewline
\midrule
Na\textsuperscript{+} & 0.838696 & 1.158723\tabularnewline
K\textsuperscript{+} & 0.682554 & 1.658967\tabularnewline
Rb\textsuperscript{+} & 0.351811 & 1.887650\tabularnewline
Cs\textsuperscript{+} & 0.374979 & 2.136627\tabularnewline
Ag\textsuperscript{+} & 0.379885 & 0.900769\tabularnewline
Cu\textsuperscript{+} & 0.300000 & 0.600000\tabularnewline
Li\textsuperscript{+} & 0.352789 & 0.735018\tabularnewline
F\textsuperscript{-} & 0.010022 & 2.061777\tabularnewline
Cl\textsuperscript{-} & 0.011348 & 2.634214\tabularnewline
Br\textsuperscript{-} & 0.010006 & 2.820870\tabularnewline
I\textsuperscript{-} & 0.013436 & 3.049587\tabularnewline
\bottomrule
\end{tabular}
\end{table}

\begin{figure*}
\includegraphics{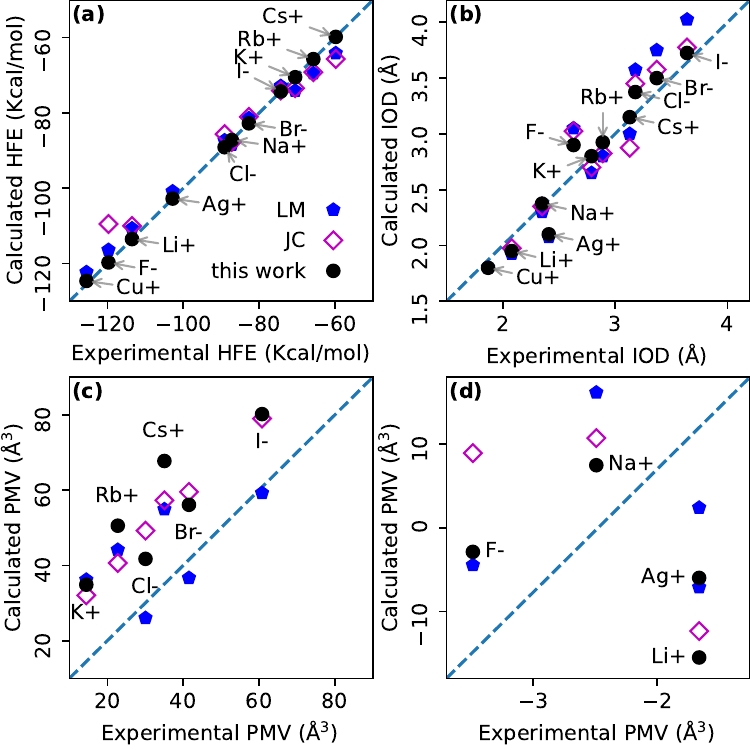}

\caption{Correlation of (a) HFE, (b) IOD and (c, d) PMV between experimental\citep{li2015systematic,couture1956partial}
and calculated results for all parameter sets. A detailed view of
ions with negative PMVs is shown in (d).\label{fig:hfe-iod-pmv}}
\end{figure*}

The new parameters lead to excellent agreement with experimental HFEs,
and IOD values\citep{li2015systematic} are improved for all ions
except Cu\textsuperscript{+}, Li\textsuperscript{+} and Na\textsuperscript{+}
relative to the LM and JC parameters (Figure \ref{fig:hfe-iod-pmv}a,b,
Supplementary Material Section S4). The largest IOD errors are observed
for F\textsuperscript{-} ($0.27$ $\text{Å}$) and Ag\textsuperscript{+}
($0.3$ $\text{Å}$), but this is still an improvement compared to
the prior parameter sets.

The weights for PMV error simply ensure rational values for this property
and values from our optimized parameter set are of similar quality
to the LM and JC parameter sets (Figure \ref{fig:hfe-iod-pmv}c,d,
Supplementary Material Section S4). All parameter sets overestimate
the PMV for positive values by a similar amount and with the exception
of F\textsuperscript{-}, all parameter sets struggled to predict
negative PMV values. The LM HFE parameters have the best PMV predictions
for Cl\textsuperscript{-}, Br\textsuperscript{-} and I\textsuperscript{-},
but also have considerably worse IOD values, which demonstrates the
tradeoff mentioned previously. As for the ions with negative PMV,
with exception of F\textsuperscript{-}, all parameter sets struggled
predicting this property. However, the errors are about the same magnitude
as for the positive PMV values.

Complete results for all models can be found in Supplementary Material
Section S4.

The values of root-mean-squared deviation and $R^{2}$ are given in
Tables \ref{tab:rmsd-r2-hfe}, \ref{tab:rmsd-r2-iod}, and \ref{tab:rmsd-r2-pmv},
which also include data for calculations using the TIP3P water model,
giving a quantitative measure of the improvements discussed previously,
and showing that, by comparing the 5th and 95th percentiles, the new
PMV predictions are comparable to most of the existing parameter sets.
Values for $R^{2}$, the mean, 5th, and 95th percentiles were calculated
via bootstrap method with 10,000 re-samplings. It is possible to see
that, for both water models, there is no parameter set with the best
performance for all three properties at the same time. For example,
LM \textcolor{black}{HFE parameters for TIP3P and} SPCE models have
the best results for PMV, but the worst results for IOD. \textcolor{black}{Furthermore,
we note that the LM 12-6-4 models had overall similar performance
to the LM HFE and JC models. }Overall, our parameter set has the best
agreement with experiment for the HFE and IOD, and average performance
for PMV, which is the balance we were aiming to achieve.

The target values that we have chosen for hydration free energies
for individual ions are not simply ``experimental'' values, since
the division of the latter between cations and anions depends upon
some extrathermodynamic assumptions that have been extensively discussed
in the literature.\citep{Marcus91,Lamoureux06a,Hunenberger11} Varying
choices are reflected in the chosen values for the HFE of the solvated
proton, which range from about $-250$ to $\unit[-264]{kcal/mol}$.\citep{Marcus91,Tissandier98,Kelly06b}
A significant contribution to this range revolves around the question
of whether to include a contribution from the phase potential for
transferring an ion across the vacuum/water interface, and if so,
what value to assign to it.\citep{Lamoureux06a,Hofer18} The Amber,
CHARMM, and AMOEBA communities have chosen values near $\unit[-251]{kcal/mol}$
based on microscopic calculations and the fact that, in periodic boundary
simulations, there is never a gas-liquid interface.\citep{Grossfield03,Lamoureux06a,Sengupta21}
We have followed their lead here, since RISM calculations are most
often used in the context of these existing empirical force fields.
It is worth noting that other microscopic calculations have supported
the more negative reference energy;\citep{Hofer18} if a more negative
proton HFE were chosen, such as $\unit[-264]{kcal/mol}$,\citep{Tissandier98}
the optimized parameters would be different from those reported here,
with cationic HFEs about $\unit[13]{kcal/mol}$ more negative and
anionic HFEs about the same amount less negative. The details of the
methods presented here should aid in any future project to derive
non-bonded parameters with a different absolute reference.

\begin{table}[H]
\caption{Hydration free energy root-mean-square deviation (kcal/mol) and $R^{2}$
values for existing parameter sets compared with experimental data\citep{li2015systematic},
including TIP3P water and and those acquired in this work. Confidence
intervals calculated via bootstrap resampling. \label{tab:rmsd-r2-hfe}}

\begin{tabular}{ccccc}
\toprule 
 & \multirow{2}{*}{RMSD} & \multicolumn{3}{c}{R\textsuperscript{2}}\tabularnewline
 &  & mean & 5th percentile & 95th percentile\tabularnewline
\midrule
LM SPC/E & 2.83 & 0.98 & 0.97 & 0.99\tabularnewline
LM TIP3P & 2.71 & 0.98 & 0.96 & 0.99\tabularnewline
JC SPC/E & 4.56 & 0.93 & 0.89 & 0.97\tabularnewline
JC TIP3P & 3.79 & 0.95 & 0.90 & 0.98\tabularnewline
New parameters & 0.27 & 1.00 & 1.00 & 1.00\tabularnewline
\bottomrule
\end{tabular}
\end{table}

\begin{table}[H]
\caption{Ion-oxygen distance root-mean-square deviation (Å) and $R^{2}$ values
for existing parameter sets compared with experimental data\citep{li2015systematic},
including TIP3P water and and those acquired in this work. Confidence
intervals calculated via bootstrap resampling. \label{tab:rmsd-r2-iod}}

\begin{tabular}{ccccc}
\toprule 
 & \multirow{2}{*}{RMSD} & \multicolumn{3}{c}{R\textsuperscript{2}}\tabularnewline
 &  & mean & 5th percentile & 95th percentile\tabularnewline
\midrule
LM SPC/E & 0.27 & 0.66 & 0.34 & 0.85\tabularnewline
LM TIP3P & 0.24 & 0.76 & 0.45 & 0.94\tabularnewline
JC SPC/E & 0.20 & 0.72 & 0.31 & 0.93\tabularnewline
JC TIP3P & 0.17 & 0.76 & 0.39 & 0.94\tabularnewline
New parameters & 0.15 & 0.89 & 0.75 & 0.97\tabularnewline
\bottomrule
\end{tabular}
\end{table}

\begin{table}[H]
\caption{Partial molar volume root-mean-square deviation (Å\protect\textsuperscript{3})
and $R^{2}$ values for existing parameter sets compared with experimental
data\citep{couture1956partial}, including TIP3P water and and those
acquired in this work. Confidence intervals calculated via bootstrap
resampling. \label{tab:rmsd-r2-pmv}}

\begin{tabular}{ccccc}
\toprule 
 & \multirow{2}{*}{RMSD} & \multicolumn{3}{c}{R\textsuperscript{2}}\tabularnewline
 &  & mean & 5th percentile & 95th percentile\tabularnewline
\midrule
LM SPC/E & 13.28 & 0.42 & -0.31 & 0.86\tabularnewline
LM TIP3P & 13.26 & -0.81 & -2.48 & 0.83\tabularnewline
JC SPC/E & 16.96 & 0.12 & -0.74 & 0.60\tabularnewline
JC TIP3P & 16.35 & 0.11 & -0.72 & 0.62\tabularnewline
New parameters & 18.14 & 0.02 & -1.05 & 0.66\tabularnewline
\bottomrule
\end{tabular}
\end{table}

\subsection{Parameter optimization: finite concentrations}

\begin{figure}
\includegraphics{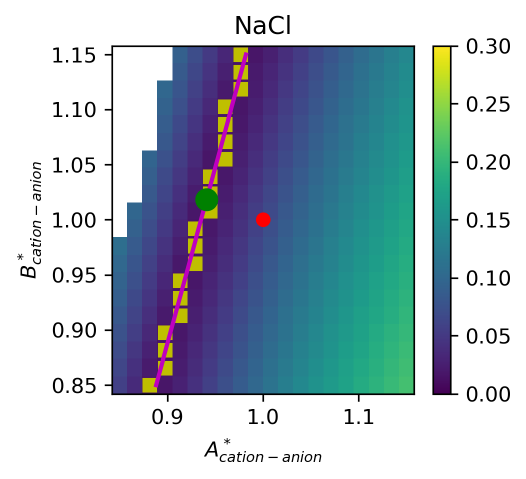}

\caption{Average relative error of the mean activity coefficient of NaCl salt
for the relative change in $A_{\text{cation\textendash anion}}$ and
$B_{\text{cation\textendash anion}}$. The colormap represents the
RMSE of the activity coefficient relative to experimental data \citep{haynes2014crc}.
The pink line is the linear fit to the lowest RMSE for each $B_{\text{cation\textendash anion}}$
value, the green circle indicates the best achievable parameters obtained
from the interpolation procedure, and the red circle denotes the Lorentz--Berthelot
(infinite dilution) parameters.\label{fig:avg-error-mac-nacl}}
\end{figure}

To assess the performance of our optimized parameters in Table \ref{tab:parameters},
we calculated the mean activity coefficient at finite concentrations
for all anion-cation monovalent ion pairs available in the LM and
JC parameter sets. As we observed for the LM and JC parameter sets,
agreement with experiment was inconsistent. Some ion pairs agreed
well with experiment, and others did not, and no parameter set was
clearly better than any other. To address this, we introduced NBFIX
parameters between the cations and anions by modifying the A\textsubscript{cation-anion}
and B\textsubscript{cation-anion} values from those obtained using
the Lorentz-Berthelot mixing rules. Figure \ref{fig:avg-error-mac-nacl}
shows representative results for NaCl, in which there is a band of
NBFIX parameters that give the lowest mean activity coefficient errors.
In most cases, the parameters derived from the mixing rules gave poor
results but were near the minimum-error band. To obtain improved activities
up to 1 M, we selected the NBFIX parameters for each ion pair as the
point from the minimum error band closest to the parameters determined
from the mixing rules as described in Section \ref{subsec:Non-bond-fix-for}.
The resulting parameters are presented in Table \ref{tab:LB} as $\epsilon$
and $\nicefrac{R_{\text{min}}}{2}$.

Except for LiCl, CsF, and RbCl, all ion pairs exhibited larger $\epsilon$
and smaller $\nicefrac{R_{\text{min}}}{2}$ parameters at finite concentration
than those given by the Lorentz--Berthelot mixing rules (Tables~\ref{tab:LB}
and~\ref{tab:nbf}). Both LiCl and CsF exhibited convergence issues
when using the Lorentz--Berthelot values, requiring shifts to smaller
$\epsilon$ and larger $\nicefrac{R_{\text{min}}}{2}$ NBFIX values
(Section S4.A). In contrast, RbCl showed no convergence problems;
its optimal parameters favored smaller $\epsilon$ and larger $\nicefrac{R_{\text{min}}}{2}$
NBFIX values than the Lorentz--Berthelot values.

Although the changes in $\epsilon$ and $\nicefrac{R_{\text{min}}}{2}$
are small, they result in stronger cation--anion interactions, except
in the cases of LiCl, CsF, and RbCl. A reduction in $\nicefrac{R_{\text{min}}}{2}$
decreases the minimum cation--anion separation, thereby strengthening
electrostatic interactions. Similarly, larger values of $\epsilon$
slightly increase the strength of the Lennard--Jones interaction
for these ion pairs.

As the NBFIX parameters are close to those calculated from the mixing
rules, NBFIX parameters could be avoided by including activities along
with infinite dilution properties as part of the initial optimization
procedure. However, this would be a much more complex optimization
problem, as all ions would need to be considered at once.

\begin{table*}
\caption{Parameters for cation-anion interactions calculated from infinite
dilution optimization using Lorentz/Berthelot mixing rules .\label{tab:LB}}

\centering{}%
\begin{tabular}{ccccccccc}
\hline 
\multirow{2}{*}{} & \multicolumn{2}{c}{Cl} & \multicolumn{2}{c}{I} & \multicolumn{2}{c}{Br} & \multicolumn{2}{c}{F}\tabularnewline
\cline{2-9}
 & $\epsilon$ (kcal/mol) & $\nicefrac{R_{min}}{2}$ (Å) & $\epsilon$ (kcal/mol) & $\nicefrac{R_{min}}{2}$ (Å) & $\epsilon$ (kcal/mol) & $\nicefrac{R_{min}}{2}$ (Å) & $\epsilon$ (kcal/mol) & $\nicefrac{R_{min}}{2}$ (Å)\tabularnewline
\hline 
\hline 
Na & 0.097556 & 3.792938 & 0.106153 & 4.208310 & 0.091606 & 3.979593 & - & -\tabularnewline
K & 0.088008 & 4.293182 & 0.095763 & 4.708555 & 0.082640 & 4.479838 & - & -\tabularnewline
Cs & 0.065231 & 4.770841 & 0.070980 & 5.186214 & 0.061253 & 4.957497 & 0.061304 & 4.198404\tabularnewline
Li & 0.063272 & 3.369232 & 0.068847 & 3.784605 & 0.059413 & 3.555888 & - & -\tabularnewline
Rb & 0.063184 & 4.521864 & 0.068752 & 4.937237 & 0.059330 & 4.708520 & - & -\tabularnewline
\hline 
\end{tabular}
\end{table*}

The final NBFIX parameters, converted to $\epsilon$ and $\nicefrac{R_{min}}{2}$,
are presented in Table \ref{tab:nbf}. Comparing the Lorentz-Berthelot
mixing rules with the acquired parameters, it is observed that there
is no large deviations, reinforcing the possibility of carrying out
an optimization for both infinite dilution and finite concentration
simultaneously.

\begin{figure}
\includegraphics{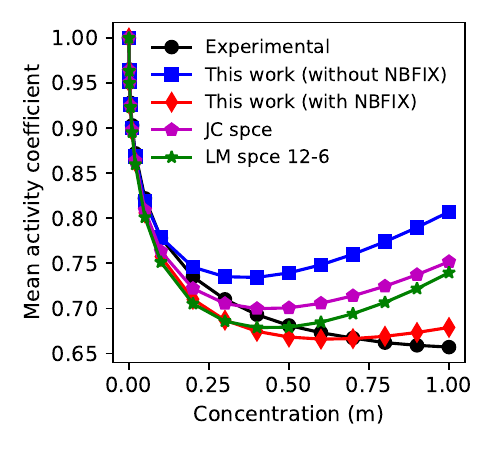}

\caption{Comparison of mean activity coefficient for NaCl from experiment\citep{haynes2014crc}
and all parameter sets with 1D-RISM.\label{fig:comparison-mac}}
\end{figure}

\begin{table*}
\caption{Parameters for cation-anion interactions acquired after nonbond fix
optimization.\label{tab:nbf}}

\centering{}%
\begin{tabular}{ccccccccc}
\hline 
\multirow{2}{*}{} & \multicolumn{2}{c}{Cl} & \multicolumn{2}{c}{I} & \multicolumn{2}{c}{Br} & \multicolumn{2}{c}{F}\tabularnewline
\cline{2-9}
 & $\epsilon$ (kcal/mol) & $\nicefrac{R_{min}}{2}$ (Å) & $\epsilon$ (kcal/mol) & $\nicefrac{R_{min}}{2}$ (Å) & $\epsilon$ (kcal/mol) & $\nicefrac{R_{min}}{2}$ (Å) & $\epsilon$ (kcal/mol) & $\nicefrac{R_{min}}{2}$ (Å)\tabularnewline
\hline 
\hline 
Na & 0.107529 & 3.743291 & 0.161291 & 3.969675 & 0.114783 & 3.857957 & - & -\tabularnewline
K & 0.095420 & 4.246648 & 0.121628 & 4.558633 & 0.095811 & 4.390537 & - & -\tabularnewline
Cs & 0.067345 & 4.750227 & 0.074339 & 5.154113 & 0.061253 & 4.957497 & 0.052394 & 4.289444\tabularnewline
Li & 0.055570 & 3.428430 & 0.107174 & 3.548912 & 0.071112 & 3.466251 & - & -\tabularnewline
Rb & 0.061689 & 4.536651 & 0.072770 & 4.899481 & 0.060151 & 4.699676 & - & -\tabularnewline
\hline 
\end{tabular}
\end{table*}
The mean activity coefficient calculated for NaCl, using the SPCE
water model, with the parameters acquired in this work using just
the mixing rules and with the NBFIX values is compared to the LM and
JC parameter sets and experimental data\citep{haynes2014crc} in Figure
\ref{fig:comparison-mac}. In this case, LM, JC, and our optimized
parameter set with just the mixing rules all diverged from the experimental
values as the concentration increased. However, the parameter set
with the NBFIX predicts mean activity coefficients that are in close
agreement with the experimental data. It is important to note that
since the changes were made only to cation-anion interactions, this
second optimization step does not change the results at infinite dilution.
Furthermore, these NBFIX parameters proposed here should only be used
in combination with our optimized parameters in Table \ref{tab:parameters}.
NBFIX parameters would need to be optimized separately for the LM
and JC parameter sets.

As observed in Table \ref{tab:rmse-mac}, the existing parameter sets
perform inconsistently across ion pairs, while our NBFIX parameters
have root-mean-squared errors (RMSEs) of less than 0.09 for all ion
pairs except CsI. In contrast, the majority of ion pairs exceed an
RMSE of 0.09 in the other models. Of the 16 salts analyzed, our optimized
parameters gave the lowest RMSE values for all but four. Therefore,
these new parameters provide consistently better results for both
infinite dilution and finite concentrations.

Complete results for all models can be found in Supplementary Material
Section S4.

\begin{table*}
\caption{Root mean squared error of the experimental\citep{haynes2014crc}
and calculated mean activity coefficient at all concentrations for
each salt. Bold values indicate best result in that row.\label{tab:rmse-mac}}

\centering{}%
\begin{tabular}{cccccc}
\toprule 
Salt & LM/SPCE & LM/TIP3P & JC/SPCE & JC/TIP3P & This work with NBFIX\tabularnewline
\midrule
NaCl & 0.0314 & 0.0691 & 0.0371 & 0.1028 & \textbf{0.0138}\tabularnewline
KCl & 0.1197 & 0.0695 & 0.0401 & 0.0764 & \textbf{0.0172}\tabularnewline
CsCl & 0.2083 & 0.1670 & 0.0860 & 0.0702 & \textbf{0.0476}\tabularnewline
LiCl & 0.1243 & \textbf{0.0808} & 0.1118 & 0.0993 & 0.1074\tabularnewline
RbCl & 0.1524 & 0.1126 & 0.0439 & 0.0216 & \textbf{0.0204}\tabularnewline
NaI & 0.5401 & 0.3588 & 0.2886 & 0.0450 & \textbf{0.0093}\tabularnewline
KI & 0.5129 & 0.3918 & 0.2161 & 0.0843 & \textbf{0.0386}\tabularnewline
CsI & 0.5873 & 0.4742 & \textbf{0.0445} & 0.1997 & 0.1391\tabularnewline
LiI & 0.4878 & 0.2595 & 0.4212 & \textbf{0.0340} & 0.0497\tabularnewline
RbI & 0.5459 & 0.4421 & 0.2265 & 0.1742 & \textbf{0.0735}\tabularnewline
NaBr & 0.1306 & 0.0503 & 0.1367 & 0.0946 & \textbf{0.0113}\tabularnewline
KBr & 0.1915 & 0.1438 & 0.1195 & 0.0234 & \textbf{0.0151}\tabularnewline
CsBr & 0.2676 & 0.2286 & 0.0909 & 0.1376 & \textbf{0.0838}\tabularnewline
LiBr & 0.0403 & 0.1196 & 0.2554 & 0.1013 & \textbf{0.0387}\tabularnewline
RbBr & 0.2216 & 0.1850 & 0.1262 & 0.0925 & \textbf{0.0182}\tabularnewline
CsF & 0.0440 & \textbf{0.0304} & 0.1176 & 0.0981 & 0.0647\tabularnewline
\bottomrule
\end{tabular}
\end{table*}

\subsection{Isothermal compressibilities as a function of salt concentration}

\begin{figure*}
\includegraphics{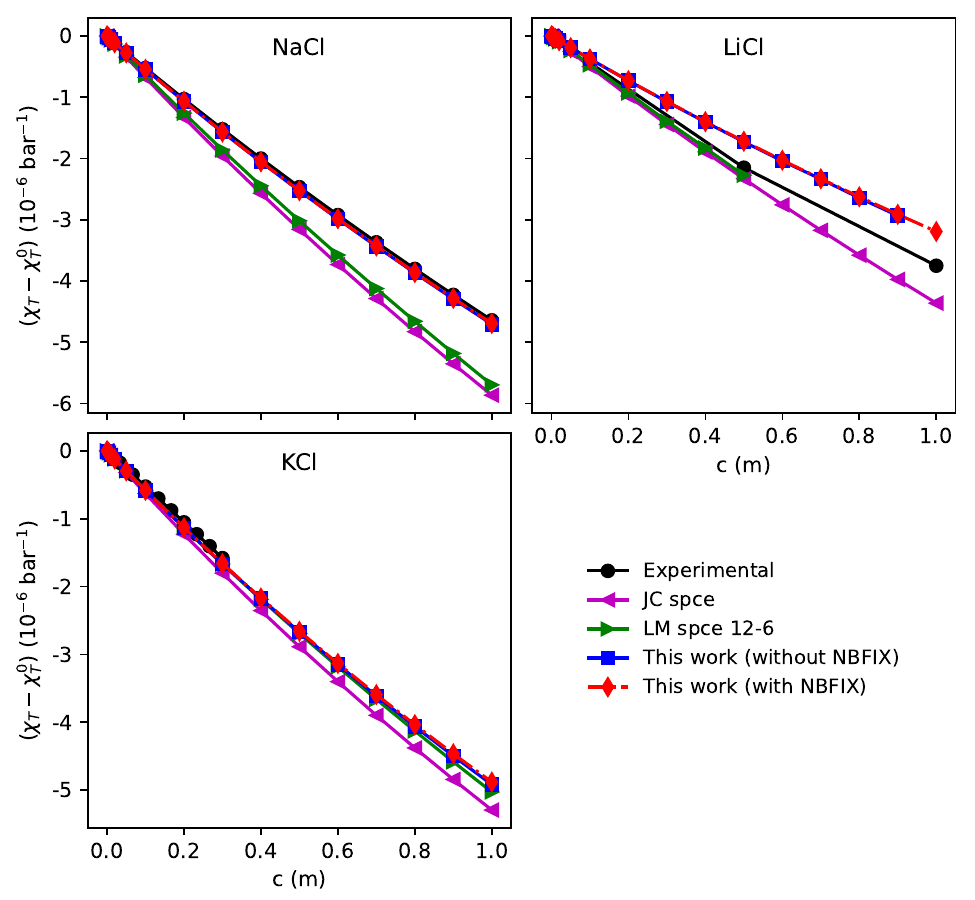}

\caption{Plot of NaCl, LiCl, KCl compressibilities from RISM and experiment.\label{fig:Plot-of-compressibilities}}
\end{figure*}

DRISM, with the PSE-3 closure relation, predicts an isothermal compressibility
for pure water, $\chi_{T}^{0}=\unit[63.46\times10^{-6}]{bar^{-1}}$,
which overestimates the experimental value of $\unit[45.25\times10^{-6}]{bar^{-1}}$\citep{vedamuthu1995properties}.
In Figure \ref{fig:Plot-of-compressibilities}, the relative compressibilities,
$\chi_{T}-\chi_{T}^{0}$, are compared with experimental data for
NaCl\citep{millero1982pvt}, LiCl\citep{apelblat2007thermodynamic},
and KCl\citep{owen1957standard}. For NaCl, the experimental data
include fitting equations describing both linear and nonlinear behavior
at different temperatures. In contrast, only three data points are
available for LiCl, and for KCl, the reference provides parameters
only for the linear regime; therefore, results for this salt are shown
only up to $\unit[0.3]{m}$.

Although DRISM with the PSE-3 closure overestimates the isothermal
compressibility of pure water, the new parameter set, both with and
without NBFIX, reproduces the relative change in isothermal compressibility
with salt concentration well for NaCl and KCl. In the case of KCl,
good agreement is also obtained with the LM model. For LiCl, better
agreement at lower concentrations is achieved with the JC and LM models.

Overall, the new parameter set shows good agreement with experimental
compressibility data for the three salts examined. However, inconsistencies
may arise for other salts, as previously observed for the mean activity
coefficients. Experimental data at the same temperature and over the
same range of concentrations were not found in the literature for
additional salts, preventing a broader assessment of the model's performance.
It is also important to note that the inclusion of NBFIX parameters
does not significantly affect the compressibility results compared
with those obtained using parameters derived under infinite dilution
conditions.

\subsection{Preferential interaction parameter calculation for 24L B-DNA}

\begin{figure}
\includegraphics{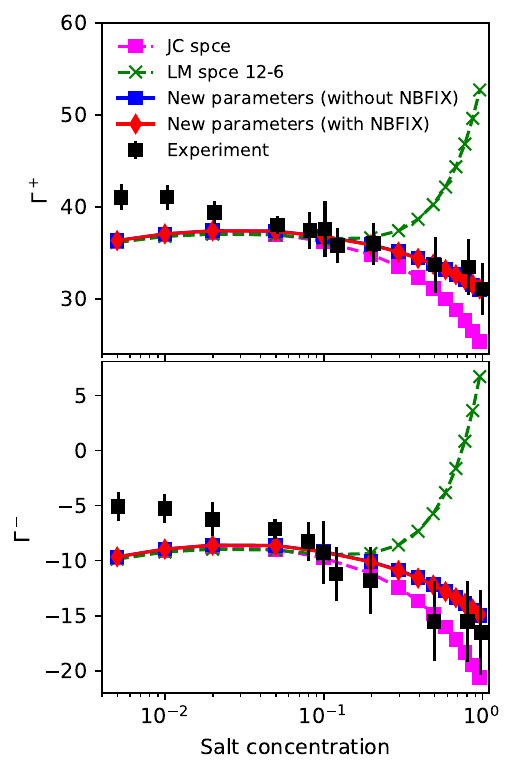}

\caption{Preferential interaction parameters for 24L nucleic acid in NaCl aqueous
solution at concentrations from $\unit[0.005]{M}$ up to $\unit[1.0]{M}$
from experiment\citep{bai2007quantitative} and calculated with 3D-RISM
using different ionic models.\label{fig:pip24L}}
\end{figure}

Our principal motivation for developing new ion parameters is to improve
the quantitative performance of predicting ion distributions around
biomolecules with 3D-RISM. As one such example, we calculated the
preferential interaction parameters (PIP) of NaCl around double stranded
DNA using the cSPC/E water model and LM, JC, and our optimized parameters
(Figure \ref{fig:pip24L}). PIP results from the new parameter set
show improved agreement with experiment at higher concentrations,
while at lower concentrations all models give essentially the same
results. Overall, we observe that improved hydration free energies
and ion-oxygen distances for the solvent acquired with the new parameters
set also lead to improved performance in 3D-RISM calculations.

\begin{figure}
\includegraphics{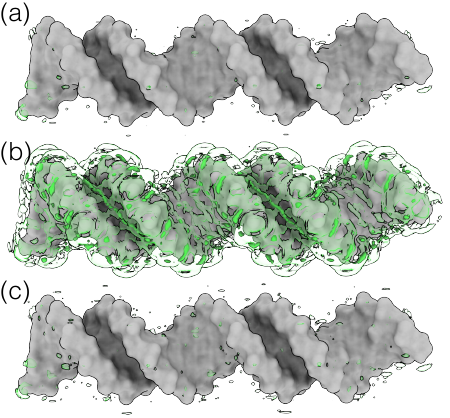}

\caption{Chloride density at concentration of $\unit[1.0]{M}$ around 24L DNA
molecule \citep{bai2007quantitative,giambacsu2014ion} for (a) JC,
(b) LM, and (c) our RISM-optimized parameters. DNA solvent accessible
surface is shown in grey, chloride density is shown as transparent
green for the $g\left(r\right)=4$ isosurface and opaque green for
the $g\left(r\right)=16$ isosurface. Only the LM parameters produce
densities high enough for the $g\left(r\right)=16$ isosurface to
be visible.\label{fig:ClDensities}}
\end{figure}

We observe the greatest difference in predicted PIP values for concentrations
of $\unit[0.1]{M}$ and above. Throughout this range, our new optimized
parameters maintain good agreement with values from experiment, with
no differences observed between the Lorentz-Berthelot and NBFIX parameters,
while the JC parameters increasingly underestimate the PIP values.
However, results from the LM parameter set rapidly increase starting
from $\unit[0.1]{M}$. By inspecting the density distribution of chloride
around the 24L molecule at concentration of $\unit[1.0]{M}$ (Figure
\ref{fig:ClDensities}) we observed that the LM parameters lead to
a high concentration of chloride ions inside the minor groove, with
a maximum $g\left(r\right)=319$. For JC and our new parameters, the
maximum values of $g\left(r\right)$ are $13$ and $17$, respectively,
and no chloride accumulation in the minor groove was observed. As
the chloride $\nicefrac{R_{min}}{2}$ value for the LM parameter set
is smaller than that in JC and our optimized parameters (Figure \ref{fig:Parameter-plots}),
it seems that LM chloride anions are small enough to accumulate inside
the minor groove and, consequently, pull more sodium ions, which explains
the PIP increase. It is important to emphasize here that these observations
are specific to 3D-RISM calculations.

In contrast, PIP calculations at low concentrations depend primarily
on the the closure, with higher order PSE-n closures giving higher
PIP values \citep{giambacsu2014ion}. As the closure order increases,
the long range interactions are better described, and the PIP predictions
becomes more accurate for dilute solutions, as can be observed in
Figure S3 from reference \citep{giambacsu2014ion} supporting material.
Furthermore, the increase in predicted PIP at low concentrations in
these prior results appear to be converging as the closure order is
increased. As the HNC closure (PSE-$\infty$) is known for its good
performance describing systems with Coulomb long-range interactions,
using higher-order closures seems appropriate. However, it is increasingly
difficult to converge solutions as the closure order increases. Furthermore,
high-order closures, such as HNC, do not correctly account for strong
correlations at short distance.\citep{iyetomi1992bridge} For higher
concentrations, where the short ranged interactions are significant,
increasing the closure order linearly increases the PIP predictions.
Indeed, PSE-3 is commonly used, as it provides a good balance between
accuracy and ease of convergence, for example Refs \citep{giambacsu2015competitive,Truong2024,Misin2016,Sharma2021,Ganyecz2022}.
PSE-3 with our optimized parameters are ideal for the physiological
range of concentrations but to achieve better predictions at all concentration
range, improved closures are necessary.

\section{Conclusions}

In this work, we optimized the LJ parameters for monovalent ions for
use with RISM calculations in AmberTools and introduced \code{generateMDL},
a new command-line tool for creating input models for 1D-RISM calculations.
While using standard Amber force fields is a strength of the RISM
framework, these new ion parameters provide better agreement with
experiment compared to earlier LM and JC parameter sets designed for
molecular dynamics. We found that the LM and JC parameter sets provided
good overall agreement with experimental HFE and IOD; however, each
set includes specific ions for which the predictions deviated substantially
from experiment. Thus, we carried out the optimization of monovalent
ions against hydration free energy, ion-oxygen distance, and partial
molar volume, calculated at infinite dilution. However, this did not
improve mean activity coefficients at finite salt concentrations,
so we introduced and optimized NBFIX parameters to account for cation-anion
interactions.

This new set of LJ parameters accurately predicts the hydration free
energies for all ions and shows overall improvement for ion-oxygen
distances as well. Although the partial molar volume could not be
optimized with the same level of accuracy, its deviation from experimental
values remains comparable to that of previously available parameter
sets. The NBFIX at finite concentrations enabled us to acquire the
best mean activity coefficient predictions for 12 out of the 16 ion
pairs analyzed. Therefore, the new set of parameters demonstrates
better consistency at finite concentrations compared to the other
existing parameter sets. Preferential interaction parameter calculations
for 24L in aqueous NaCl solution using the new set of parameters also
demonstrate improvements at higher concentrations compared to the
existing parameter sets. We observed that calculations using LM parameters
lead to unexpected results at higher concentrations, possibly due
to the smaller chloride ion size, which caused RISM to predict a higher
concentration of this ion in the DNA minor groove. Comparison with
experimental isothermal compressibility data also shows good agreement
for the relative change with respect to pure water for the new set
of parameters. For both isothermal compressibility and PIP calculations,
no differences were observed between the NBFIX parameters and the
original parameters obtained under infinite dilution conditions. However,
the NBFIX parameters may be of practical utility since numerically
stability at higher concentrations was part of the selection criteria.

The results presented in this work show that the new set of parameters
developed for the RISM framework represents an improvement over the
existing parameter sets for 1) 1D-RISM calculations at infinite dilution
and finite salt concentrations, 2) 1D-RISM predictions of relative
isothermal compressibilities for NaCl and KCl, and 3) 3D-RISM prediction
of experimentally measured PIPs for a DNA solute. However, we observed
no improvement for PIP calculations at low concentrations, which is
likely a limitation of the closure used. Therefore, further development
of improved closures remains necessary.

\section*{SUPPLEMENTARY MATERIAL}

See the Supplementary Material for a derivation of Equation (\ref{eq:mean-activity}),
a discussion of the thermodynamic consistency of the solvent-solvent
and solute-solvent routes, an analysis of the influence of the cost
function weights on the optimized parameters, a description of how
convergence issues were handled during the finite-concentration optimization
step, and a table of dielectric constants for all salts at 1 M concentration.
In addition, raw data for calculated and experimental HFE, IOD, PMV,
mean activities, and isothermal compressibilities are provided. Complete
scripts and input files to reproduce all data are available on Zenodo.\citep{carvalho2026zenodo}
\begin{acknowledgments}
This material is based upon work supported by the National Science
Foundation (NSF) under Grants CHE-2102668, CHE-2018427, CHE-2320718
and MRI-2320846.
\end{acknowledgments}

\section*{AUTHOR DECLARATIONS}

\subsection*{Conflict of Interest}

The authors have no conflicts to disclose.

\subsection*{Author Contributions}

\strong{Felipe Silva Carvalho}: Conceptualization (equal); Data Curation
(equal); Formal Analysis (equal); Investigation (equal); Methodology
(equal); Software (equal); Visualization (equal); Writing/Original
Draft Preparation (equal); Writing/Review \& Editing (equal). \strong{Alexander McMahon}:
Conceptualization (equal); Data Curation (equal); Formal Analysis
(equal); Investigation (equal); Methodology (equal); Software (equal);
Visualization (equal); Writing/Original Draft Preparation (equal);
Writing/Review \& Editing (equal). \strong{David A. Case}: Conceptualization
(equal); Funding Acquisition (equal); Methodology (equal); Visualization
(equal); Writing/Original Draft Preparation (equal); Writing/Review
\& Editing (equal). \strong{Tyler Luchko}: Conceptualization (equal);
Data Curation (equal); Funding Acquisition (equal); Methodology (equal);
Project Administration (equal); Visualization (equal); Writing/Original
Draft Preparation (equal); Writing/Review \& Editing (equal).

\section*{DATA AVAILABILITY}

The data that support the findings of this study are available within
this article and its Supplementary Material. Scripts and input files
required to reproduce the calculations with AmberTools 26 have been
deposited in Zenodo at DOI: 10.5281/zenodo.19211784.\citep{carvalho2026zenodo}

\bibliography{manuscript}

\end{document}